\documentclass{emulateapj}

\usepackage{hyperref}
\usepackage{xcolor}
\usepackage{url}
\usepackage{amssymb}
\usepackage{physics}
\hypersetup{
    colorlinks,
    linkcolor={red!50!black},
    citecolor={blue!50!black},
    urlcolor={black} 
}
\usepackage{mathrsfs}
\usepackage{nicefrac}

\shorttitle{NIRSPEC Spectra of NLTT 5306}
\shortauthors{Buzard et al.}

\begin{document}

\title{Near-infrared spectra of the inflated post-common envelope brown dwarf NLTT 5306B}

\author{Cam Buzard\altaffilmark{1}, Sarah Casewell\altaffilmark{2}, Joshua Lothringer\altaffilmark{3,4}, Geoffrey A. Blake\altaffilmark{1,5}}

\altaffiltext{1}{Division of Chemistry and Chemical Engineering, California Institute of Technology, Pasadena, CA 91125, USA}
\altaffiltext{2}{School of Physics and Astronomy, University of Leicester, University Road, Leicester LE1 7RH, UK}
\altaffiltext{3}{Department of Physics and Astronomy, Johns Hopkins University, Baltimore, MD, USA}
\altaffiltext{4}{Department of Earth \& Planetary Sciences, Johns Hopkins University, Baltimore, MD, USA}
\altaffiltext{5}{Division of Geological and Planetary Sciences, California Institute of Technology, Pasadena, CA 91125, USA}

\begin{abstract}
NLTT 5306 is a post-common envelope binary made up of a white dwarf host and brown dwarf companion that has shown evidence of inflation and active mass donation despite not filling its Roche lobe. Two proposed mechanisms for the brown dwarf's inflation are magnetic interactions and a high metallicity, cloudy atmosphere. We present moderate resolution ($R\lesssim2000$) $J$ band Keck/NIRSPEC observations of this system. These phase-resolved data allow us to constrain differences between atmospheric parameters of the day- and night-side of the brown dwarf. Our day- and night-side effective temperature measurements are consistent, in agreement with the brightness temperatures measurements from \citealt{Casewell2020nltt5306}. The day-side favors a slightly lower surface gravity, perhaps stemming from the material streaming between the two objects. Finally, our data show a preference for low metallicity models. This would be expected from the system's old age, but provides direct evidence that a high metallicity, cloudy brown dwarf atmosphere is not responsible for the witnessed inflation. These results strengthen the case for magnetic interactions leading to inflation of NLTT 5306 B.
\end{abstract}

\section{Introduction}

As exoplanet detecting surveys ramped up the number of known planets orbiting other stars, an interesting phenomenon arose. The ``brown dwarf desert" describes the apparent lack of brown dwarf ($\sim 10-80\ M_{Jup}$) companions within $\sim$5 AU of solar-type stars \citep[e.g.,][]{Grether2006}. \citealt{Grieves2017} estimated the brown dwarf occurrence rate around solar-type stars with periods less than 300 days to be $\sim$0.56\%. It follows that the evolved form of these rare binaries, white dwarf/brown dwarf binaries, are also quite rare; \citealt{Steele2011} predicted only $0.5 \pm 0.3$\% of white dwarfs have brown dwarf companions. To date, only eleven detached systems are known \citep{vanRoestel2021} and the number of interacting systems is equally small \citep[e.g.,][]{Burleigh06,hernandez16}. 


White dwarf/brown dwarf binaries are often called substellar post-common envelope binaries because of the chaotic evolutionary pathway they go through as their host star dies and evolves into a white dwarf. When the host star expands once it has exhausted its hydrogen reservoir, it overfills its Roche Lobe and eventually enters a common envelope with its brown dwarf companion. During this stage, the brown dwarf's orbit becomes unstable and it begins to spiral inward. The dying star loses its outer layers and leaves its core as a white dwarf, ending the common envelope period and retreating back within its Roche Lobe. \citealt{Izzard2012} offers a nice review of the common envelope process. After the common envelope has dissipated, in the ``post-common envelope binary" stage, the brown dwarf is still on an inspiraling orbit and eventually it will fill its Roche Lobe and begin to donate mass to the white dwarf. SDSS~J121209.31+013627.7 \citep{Burleigh06, Stelzer17} and SDSS~J143317.78+101123.3 \citep{hernandez16} are two examples of systems at this stage in the evolutionary process, both showing mass donation from the brown dwarf companion onto the white dwarf host.  

NLTT 5306 is a post-common envelope binary with a brown dwarf companion. Details of this system are given in Table~\ref{NLTTproperties}. It was first discovered by \citet{Steele2013} who searched for known white dwarfs with an infrared excess indicative of a cool companion, and it was initially thought to be a detached system. However, evidence in the form of a weak hydrogen emission feature that moves in phase with and at the radial velocity of the white dwarf and not the brown dwarf, and a sodium absorption feature moving the same way, suggests that the brown dwarf may have just begun losing mass to its host white dwarf \citep{longstaff2019}. Non-detections in both X-rays and the radio at 6 GHz put an upper limit of this accretion at 1.3$\times 10^{11}$~gs$^{-1}$ \citep{longstaff2019}. This low accretion rate, and the geometry of the system, suggests the brown dwarf is not, in fact, filling its Roche Lobe, and as such, the mechanism leading to accretion is unclear.

While NLTT 5306 B does not fill its Roche Lobe, near-IR spectra from SpeX on IRTF have shown evidence of intermediate gravity, with $\log(g)\sim$4.8, suggesting the brown dwarf is inflated \citep{Casewell2020nltt5306}. In exoplanets, significant levels of ultraviolet (UV) irradiation onto a planet from its host can cause the planetary atmosphere to inflate and evaporate \citep[e.g.,][]{Demory2011}. The white dwarf, NLTT 5306 A, however, is relatively cool ($T_{\mathrm{eff}}=7756 \pm 35$ K), and so it is not possible that the brown dwarf is having its atmosphere ``boiled off." In fact, there are brown dwarfs in closer orbits around hotter white dwarfs which show no evidence of either inflation or mass accretion onto the white dwarf. Two such examples are SDSS J1205-0242B \citep{Parsons2017} and SDSS J1411+2009B \citep{Littlefair2014}, which receive $\sim$250 and $\sim$4.5$\times$ the irradiation of NLTT 5306B, respectively. On the other hand, NGTS-19b, a high mass brown dwarf orbiting a K dwarf \citep{Acton2021}, and WD1032+011, a brown dwarf orbiting a 9950 K white dwarf \citep{Casewell2020WD1032}, like NLTT 5306B, both receive relatively little UV irradiation and yet, are both inflated. 

Inflation and low surface gravity have long been understood to be indicators of youth in brown dwarfs \citep[e.g.,][]{Cruz2009,Allers2013}. When under $\sim100$ Myr, brown dwarfs are still contracting and so have larger radii than older brown dwarfs of the same spectral type \citep{Burrows2001}. \citealt{Casewell2020nltt5306} measured the gravity sensitive indices of \citealt{Allers2013} from their SpeX spectrum of NLTT 5306B and found that its intermediate gravity was comparable to that of a $50-200$ Myr L5 brown dwarf. NLTT 5306B is known to be much older than that though. As a known thick disk object \citep{Steele2013}, NLTT 5306B is at least 5 Gyr (its white dwarf's minimum cooling age) and probably much older. Youth is not a plausible explanation of the inflation seen in this dwarf.  

\citealt{Sainsbury2021} considered whether a mechanism involving heating the deep atmosphere by vertical advection of potential temperature as is used to explain hot Jupiter inflation \citep{Tremblin2017,Sainsbury2019} could also explain brown dwarf inflation. They found that the inflation of brown dwarfs Kepler-13Ab and KELT-1b, orbiting 7650 and 6518 K main-sequence stars, could be explained this way. The highly irradiated 13000 K white dwarf companion SDSS1411B could not be inflated by vertical advection of energy into its deep atmosphere. \citealt{Sainsbury2021} suggest that the ineffectiveness of vertical advection to lead to inflation of SDSS1411B could be due to its fast rotation rate. With NLTT 5306B's slightly faster rotation rate and lower irradiation, it is unclear whether heating of its deep atmosphere could lead to inflation, or not, as in the case of SDSS1411B.  


\citealt{Casewell2020nltt5306} proposed two alternate mechanisms that could lead to the inflation of NLTT 5306 B. Magnetic activity has been used to explain inflation in M dwarfs \citep{Parsons2018}. There is already some evidence that NLTT 5306 A has a non-negligible magnetic field, because there was no infrared excess indicative of an accretion disk \citep{longstaff2019}, suggesting that the accretion onto the white dwarf may be following magnetic field lines, as happens in polars. Further, when \citealt{Casewell2020nltt5306} compared NLTT 5306 B to the 23 known brown dwarfs transiting main sequence stars \citep{Carmichael2020}, they found that CoRoT-15b and CoRoT-33b, the only two orbiting magnetically active stars, were also possibly inflated. The second contributing factor they proposed was a high metallicity, cloudy brown dwarf atmosphere, which \citealt{Burrows2011} showed could lead to larger radii. NLTT 5306 B, as mid-L dwarf, could be reasonably expected to be cloudy; however, NLTT 5306 is known to be a thick disc object \citep{Steele2013}, which implies it is probably much older than its white dwarf minimum cooling age ($>5$ Gyr). Given its old age, it is unlikely to be metal-enriched.


In this work, we obtained the highest resolution near-infrared spectrum of any white dwarf/brown dwarf binary to date. With this $R\lesssim2000$ resolution spectrum of NLTT 5306, we aim to constrain the effective temperature, surface gravity, and metallicity of the brown dwarf's atmosphere and to determine whether there are any day- to night-side variations in these parameters that could provide clues into the mechanism of inflation. While evidence has pointed to weak accretion onto the white dwarf, we do not know precisely what could cause the accretion in this system. A closer picture of the brown dwarf, in the form of phase-resolved spectra, may offer some insight.

\begin{deluxetable}{lll} 
\tablewidth{0pt}
\tabletypesize{\scriptsize}
\tablecaption{NLTT 5306 System Parameters}
\tablehead{Property & Value & Ref. }
\startdata
\sidehead{\textbf{NLTT 5306A}}
Temperature & 7756 $\pm$ 35 K & (1)\\
Surface gravity, $\log(g)$ & 7.68 $\pm$ 0.08 & (1)\\
Cooling Age & 710 $\pm$ 50 Myr & (1) \\
Distance & 71 $\pm$ 4 pc & (1)\\
Mass & 0.44 $\pm$ 0.04 M$_{\sun}$ & (1)\\
Radius & 0.0156 $\pm$ 0.0016 R$_{\sun}$ & (1) \\
Systemic velocity$^{\tablenotemark{a}}$, $v_{\mathrm{sys}}$ & 15.6 $\pm$ 1.8 km/s & (2) \\ 
Velocity semi-amplitude$^{\tablenotemark{a}}$, $K_{\mathrm{WD}}$ & -46.8 $\pm$ 2.5 km/s & (2)\\
\sidehead{\textbf{NLTT 5306B}}
Orbital Period & 101.88 $\pm$ 0.02 min & (1) \\
Separation, $a$ & 0.566 $\pm$ 0.005 R$_{\sun}$ & (1) \\
Time of inferior conjunction, $T_o$ & 2453740.1778(8)  & (2) \\
Minimum Mass, $M_{\mathrm{BD}}\sin (i)$ & 56 $\pm$ 3 M$_{Jup}$ & (1) \\
Evolutionary Radius$^{\tablenotemark{b}}$ & 0.095 $\pm$ 0.004 R$_{\sun}$ & (1)\\
Roche Lobe Radius & 0.12 $\pm$ 0.02 R$_{\sun}$ & (2)\\
Spectral Type & L5 & (3)
\enddata
\label{NLTTproperties}
\tablerefs{(1) \citealt{Steele2013}, (2) \citealt{longstaff2019}, (3) \citealt{Casewell2020nltt5306}} 
\tablenotetext{a}{These values were measured from H$\alpha$ absorption lines in the white dwarf.}
\tablenotetext{b}{This radius was estimated from evolutionary models. Because the system is not eclipsing, a true radius measurement cannot be made.}
\end{deluxetable}


\section{NIRSPEC Observations and Data Reduction}

\subsection{Observations}   \label{Section:NLTTobservations}
\begin{figure}
    \centering
    \noindent\includegraphics[width=21pc]{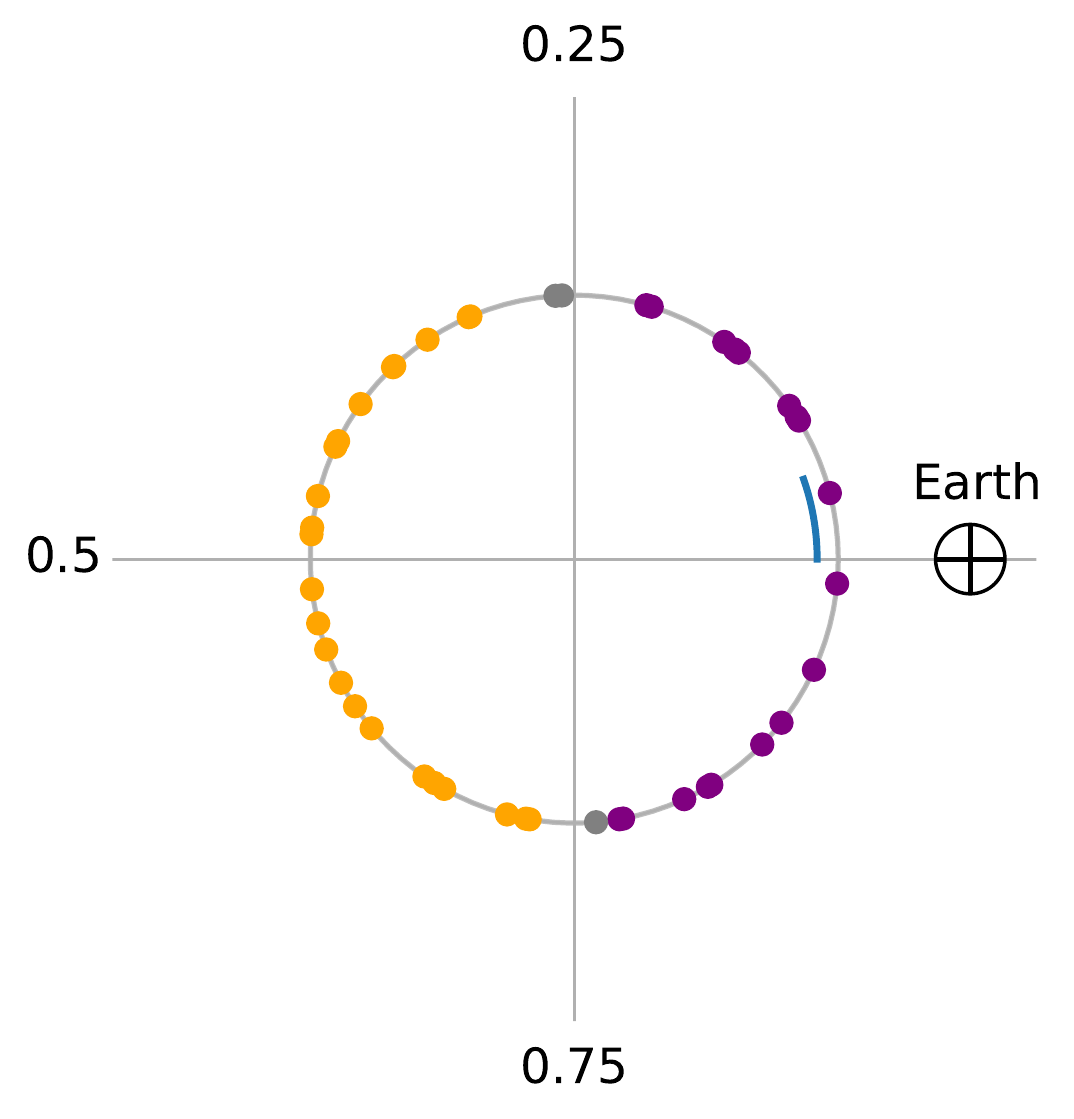}
    \caption{Diagram showing the center position of each of our 44 epochs. Orange points represent day-side epochs, purple points represent night-side epochs, and gray points represent observations during which NLTT 5306B crossed quadrature and are therefore neither primarily day- or night-side. The blue arc starting at $\phi=0$ shows the average change in orbital position over a 5 minute exposure.    }
    \label{Figure:orbit}
\end{figure}

We obtained Keck/NIRSPEC \citep{McLean1998, Martin2018} data of NLTT 5306 on three nights: October 17, 2019, January 7, 2020, and January 7, 2021. Each time, we used the NIRSPEC in its low resolution (cross disperser-only) mode with the 42$\times$0.760" slit and five minute exposures to maximize our signal-to-noise on this faint ($K_{mag}=15.6$) target. We measured a spectral resolution of $R\lesssim 2000$. On each night we observed telluric standards (HIP16322 and 31 Psc) at different airmasses to aid in telluric correcting the target data. Additional observational parameters are given in Table~\ref{nlttobservationtable}, and Figure~\ref{Figure:orbit} shows the brown dwarf orbital position at each nod. This represents the highest resolution (near-)infrared spectra ever taken of a white dwarf/brown dwarf binary system.  


\begin{deluxetable}{lccc} 
\tablewidth{0pt}
\def\arraystretch{1}
\tablecaption{NIRSPEC Observations of NLTT 5306}
\tablehead{Property & 2019 Oct 17 & 2020 Jan 7  & 2021 Jan 7  }
\startdata
Filter  & NIRSPEC-2 & NIRSPEC-1 & NIRSPEC-1 \\
Wavelength ($\mu$m)  & 1.089 -- 1.293 & 0.947 -- 1.121 & 0.947 -- 1.121 \\
Airmass  & 1.02 -- 1.17 & 1.015 -- 1.075 & 1.0 -- 1.11 \\
N$_{\mathrm{Nods}}$\tablenotemark{a}  & 14 & 10 & 20 \\
N$_{\mathrm{Nods, day}}$  & 8 & 7 & 8 \\
N$_{\mathrm{Nods, night}}$  & 5 & 3 & 10 \\
$v_{\mathrm{bary}}$ (km/s) & 1.6  & -29.6 & -29.7 
\enddata
\label{nlttobservationtable}
\tablenotetext{a}{Each nod had an exposure time of 5 minutes.}
\end{deluxetable}

\subsection{Reduction} \label{Section:Reduction}
With the two-dimensional images in hand, we flat-fielded and dark subtracted the data according to \citealt{Boogert2002}. We subtracted the A and B nods to reduce background light. Then, we used a third-order polynomial to fit and correct for any curvature of the trace, which can be quite significant in the low resolution mode of NIRSPEC, before extracting the one-dimensional spectra. 

Typically subtracting the A and B nods does not correct for telluric absorption features, which require a source of background light and so only show up, spatially separated, across the traces, but does correct for telluric emission features, which, because they do not require a source of background light, show up across the full spatial dimension of the order. Because NLTT 5306 is so faint, though, and our nods were each 5 minutes long, there was enough time for the sky emission lines to change in shape and intensity from one nod to the next. As a result, subtracting the A and B nods did not fully correct sky emission lines from our data. To compensate for this, we extracted the one-dimensional sky emission spectra off the target trace and linearly scaled them to fit the emission features in the data. We calibrated the data wavelength axes by fitting these sky emission spectra to a sky emission model from \texttt{SkyCalc}\footnote{http://www.eso.org/sci/software/pipelines/skytools/skycalc} \citep{Noll2012,Jones2013} with a fourth order polynomial fit between data pixels and model wavelengths. To correct for the sky emission lines in the data, we chose to incorporate them in our cross correlation analysis rather than divide them out. We describe how they are included in our cross correlations in Section~\ref{Section:CC}. We find that correcting the emission features this way reduces unwanted structured in cross correlation space.     

The standard data were reduced the same way as the target data. Several Paschen series hydrogen absorption features were present in the standards, including Pa-$\epsilon$ (954 nm), Pa-$\delta$ (1005 nm), and Pa-$\gamma$ (1094 nm) in the Jan 7 2020 and 2021 data taken with the NIRSPEC-1 filter and Pa-$\gamma$ (1094 nm) and Pa-$\beta$ (1282 nm) in the Oct 17 data taken with the NIRSPEC-2 filter. We masked out these features, interpolated between the airmasses of the standard observations to the airmass of each target observation, and divided each target nod by the appropriate airmass standard to remove telluric absorption features from the data. As seen in Figure~\ref{Figure:DatawithSky}, there is a jump between data from the two filters. This jump should not affect our cross correlation analysis; cross correlations are sensitive to the variation in a spectrum rather than the baseline height. As will be described in Section~\ref{Section:CC}, each nod (and so, as follows, each filter) is cross correlated separately, so the jump between filters will not be interpreted as a real absorption feature.

In order to measure the instrument profile of the data, we used the ESO tool \texttt{Molecfit} \citep{Smette2015,Kausch2015} to fit the tellurics in the standard data from each night. With the lack of telluric features present in the wavelengths covered by the Jan 7 nights, \texttt{Molecfit} could not get a good telluric fit. It was able to fit the Oct 17 data, however, and reported a Gaussian kernel for the instrumental profile with a $\sigma$ of 1.88 cm$^{-1}$. This was consistent with the kernel needed to broadened a \texttt{SkyCalc} model to fit the sky emission lines in our data from all three nights. A Gaussian kernel of 1.88 cm$^{-1}$ corresponds to R$\sim 2000$. In the later cross correlation analysis, we broaden each of our brown dwarf spectral models with this kernel before cross correlating.





\subsection{NIRSPEC Data} 

\begin{figure*}
    \centering
    \noindent\includegraphics[width=42pc]{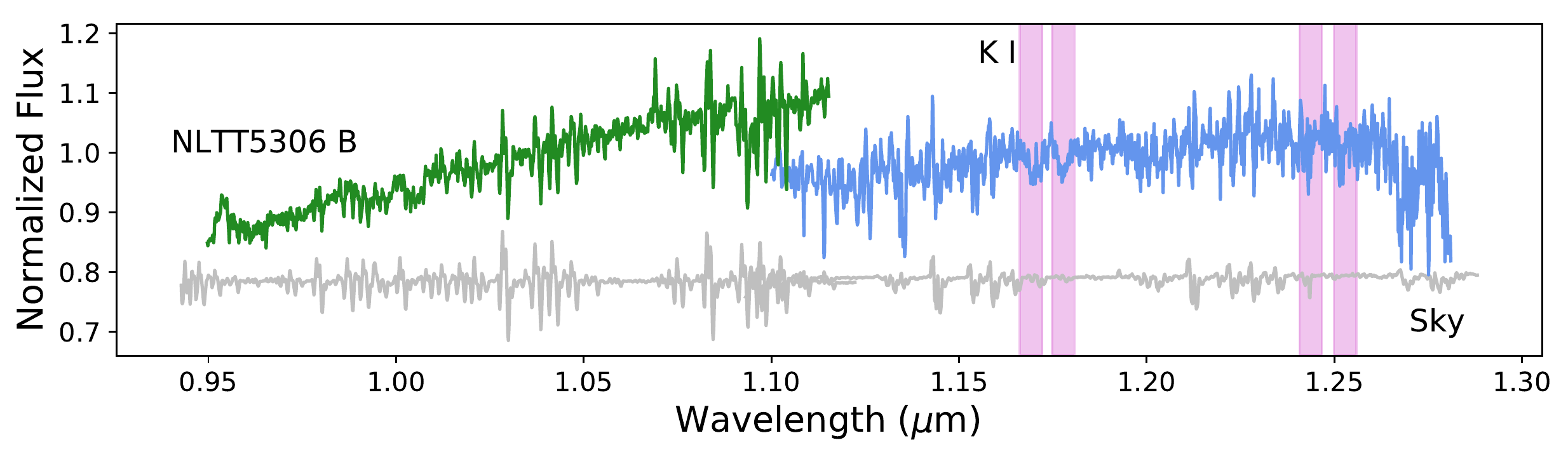}
    \caption{Our NIRSPEC data are shown in blue (Oct 17, 2019) and green (Jan 7, 2020 and Jan 7, 2021), shifted into the brown dwarf reference frame, assuming a $K_{\mathrm{BD}}$ of 333 km/s. In gray are the sky emission spectra from each epoch of data shifted in the same way as the data and vertically offset for clarity. The pink shading represent the positions of two K I doublets. Notably, the K I doublet at $\sim1.16-1.17 \mu$m is present in our data. The longer wavelength doublet coincides with a noisier portion of our spectrum.  }
    \label{Figure:DatawithSky}
\end{figure*}

The final one-dimensional spectra, shifted into the brown dwarf frame-of-reference assuming a $K_{\mathrm{BD}}$ of 333 km/s (the center of the $K_{\mathrm{BD}}$ prior described in Section~\ref{Section:KBDPrior}) and coadded, are shown in Figure~\ref{Figure:DatawithSky}. The blue and green portions of the spectrum are from the different wavelength filters. In gray, the sky emission data extracted from off of the target traces is shifted and coadded in the same way as the corresponding trace. These wavelengths fall on the Rayleigh-Jeans side of the 7756 K white dwarf blackbody curve meaning that, while the white dwarf contributes significant flux to our data, it should not add more than a linear slope to the continuum (see Figure 5 from \citealt{longstaff2019}), which is effectively removed by the standard correction. The white dwarf should show the same hydrogen absorption features we saw in the standard (Pa-$\epsilon$ at 0.954 $\mu$m, Pa-$\delta$ at 1.005 $\mu$m, Pa-$\gamma$ at 1.094 $\mu$m, Pa-$\beta$ at 1.282 $\mu$m), but we do not see strong evidence of these lines. Though, except for Pa-$\gamma$, all would fall at the noisy edges of our two orders.    

Two notable features stand out. The regions shaded in pink denote the K I doublet wavelengths. The shorter wavelength doublet appears in our data. The longer wavelength doublet is less visible, but also corresponds to a noisier portion of our data. 

Also, the gray sky emission spectrum can help us differentiate between signal from the target, noise, and telluric contamination. The $\sim$1.083 $\mu$m region where metastable He emission has been detected from exoplanets with extended, eroding atmospheres \citep[e.g.,][]{Spake2018}, is dominated by sky emission, likely OH lines, in our data. It would be interesting to see whether metastable He emission lines would arise in this brown dwarf, as it does appear to be eroding at a similar rate to the canonical WASP-107b ($\lesssim 1.3\times 10^{11}$ g/s, \citealt{longstaff2019} versus $10^{10}-3\times 10^{11}$ g/s, \citealt{Spake2018}), but as a brown dwarf, is much more dense and expresses much stronger gravitational forces on its atmosphere. Even higher resolution data, or data from a space-based instrument, would be needed to separate the telluric OH emission from any metastable He emission from the brown dwarf.

\section{Brown Dwarf Spectral Models} \label{sonoramodels}
To attempt to constrain NLTT 5306 B's effective temperature, surface gravity, and metallicity, we cross correlate our data with two sets of brown dwarf spectral models: the Sonora 2021 model grid \citep{Sonora2021} and a grid of irradiated brown dwarf models based on those presented in \citealt{Lothringer2020}. Before describing the cross correlation analysis, we wish to make a few notes on the spectral models. 



To test effective temperature, surface gravity, and metallicity, we use a subset of the Sonora 2021 model grid \citep{Sonora2021} that contains effective temperatures ($T_{\mathrm{eff}}$) from 200-600 K in steps of 25 K, 600-1000 K in steps of 50 K, and 1000-2400 K in steps of 100 K; surface gravity (log(g)) values of 3, 3.25, 3.5, 4, 4.5, 5, and 5.5; and metallicity ([Fe/H]) values of -0.5, 0, and 0.5. All of the models we use have a solar C/O ratio.  

Effective temperature has the most dramatic effect on the morphology of these spectra. From the hottest to the coldest models, absorption features from refractory species (e.g. FeH, VO, TiO) and alkali metals (Na, K) gradually dissipate, leaving spectra shaped by only H$_2$O, CH$_4$, and NH$_{3}$ below $T_{\mathrm{eff}} \approx 1000$ K \citep{Sonora2021}. Surface gravity affects the shape of the absorption features in the typical brown dwarf $J$ band, with low surface gravity objects showing weaker FeH (0.99, 1.2 $\mu$m), Na I (1.14 $\mu$m), and K I (1.17, 1.25 $\mu$m) absorption, but stronger VO (1.06 $\mu$m) absorption, than field gravity objects \citep{Allers2013}. There is some commonality in the effect of high surface gravity and low metallicity on the spectral morphology. Like high surface gravity relative to low surface gravity models, low metallicity models show stronger FeH, Na I, and K I features. However, unlike gravity, which affects both the depth and the width of the alkali features, metallicity mainly affects the depth alone. 

Additionally, the Sonora models are made to replicate the spectra of non-irradiated substellar atmospheres. \citealt{Zhou2021} recently found that the non-irradiated Sonora 2018 cloudless grid \citep{Sonora2018} resulted in poor fits to two other white dwarf/brown dwarf binaries, WD 0137B and EPIC 2122B. Irradiated models resulted in much better fits. These systems have substantially hotter white dwarfs than NLTT 5306 A, though. WD 0137A and EPIC 2122A are 16500 and 24900 K, respectively, compared to NLTT 5306A's 7756 K. It follows that WD 0137B and EPIC 2122B receive much higher levels of irradiation than NLTT 5306B. Indeed, both show H$\alpha$ and metal emission from the surface of the brown dwarf, unlike NLTT 5306 B. Nonetheless, we also consider a grid of spectral models that do include irradiation.

Our irradiated models are based on those presented in \citealt{Lothringer2020}. The model grid spans a range of internal temperatures ($T_{\mathrm{int}} = $ 1000, 1500, 2000 K), surface gravities ($\log(g) = $ 4.5, 4.75, 5.0), metallicities ([Fe/H] = -0.5, 0, 0.5), and ``irradiations." The irradiation cases include dayside heat redistribution ($f=0.5$), full planet-wide redistribution ($f=0.25$), and a high-albedo scenario ($f=0.125$). The redistribution parameter considers both the surface area over which the object cools and the albedo. Setting the redistribution to 0 implies an albedo of 1, and so describes essentially a non-irradiated object. When irradiation is removed, these models approximate the Sonora spectra. When the redistribution is non-zero, the irradiation spectrum is determined from \citealt{Koester2010} white dwarf models. It should be noted that if there is a hot spot on the white dwarf associated with its inferred mass accretion \citep{longstaff2019}, the flux the brown dwarf receives may exceed that predicted by the \citealt{Koester2010} models.

The internal temperatures and heat redistributions together determine the effective temperature of the irradiated brown dwarf. The following equations, originally from \citealt{Lothringer2020}, describe this conversion. First the irradiation temperature is determined from properties of the host and the brown dwarf heat redistribution and albedo, 
\begin{equation}  \label{Equation:Tintf_Teff}
    T_{\mathrm{irr,BD}} = (f*(1-A_{\mathrm{BD}}))^{1/4}*T_{\mathrm{eff,WD}}\sqrt{R_{\mathrm{WD}}/a}, 
\end{equation}
and then the effective temperature considers both the brown dwarf internal and irradiated temperatures,
\begin{equation}
    T_{\mathrm{eff,BD}} = (T_{\mathrm{int,BD}}^4 + T_{\mathrm{irr,BD}}^4)^{1/4}.
\end{equation}
Our three internal temperatures and three heat redistributions give rise to nine different effective temperature models. Using the white dwarf effective temperature and radius and the separation between the two objects from Table~\ref{NLTTproperties}, and a Bond albedo $A_{\mathrm{BD}}$ of 0, the combinations of internal temperatures and heat redistributions give our models effective temperatures of 1077, 1140, 1241, 1525, 1549, 1593, 2011, 2021, and 2042 K. The effective temperatures are closely grouped around the internal temperatures because the irradiation is relatively mild due to the low white dwarf effective temperature. Especially at the higher internal temperatures ($T_{int}=2000$ K), the internal temperature dominates the effective temperature, with irradiation playing only a minor role. As mentioned above, choosing an albedo of 1 removes the irradiation component and sets the effective temperature equal to the internal temperature.




\section{Cross Correlation Analysis} \label{Section:CC}

We cross correlate each of our nods with a brown dwarf model to determine the brown dwarf's line-of-sight velocity at that orbital position. The collection of velocities at different orbital positions can be used to measure NLTT 5306B's line-of-sight Keplerian orbital velocity $K_{\mathrm{BD}}$.

As described in Section~\ref{Section:Reduction}, because our exposures were long, telluric emission features made it into our data. We decided to account for them by running a two-dimensional cross correlation. The first dimension correlates the off-trace sky emission spectra with the on-trace target spectra. These should, and do, find a maximum at 0 km/s because the data are taken from the Earth's reference frame. The second dimension of the cross correlation tests a brown dwarf spectral model against the data. We find that considering the sky emission lines in this two-dimensional cross correlation framework allows us to better measure the brown dwarf velocity than by dividing out the emission features.

After each nod is cross correlated, we convert the cross correlations to log likelihoods so that they can be combined. To do so, we follow the formula presented by \citealt{Zucker2003},
\begin{equation} \label{logLequation}
    \log(L(v_{\mathrm{BD}})) = -\frac{N}{2}\log(1-C(v_{\mathrm{BD}})^2),
\end{equation}
where N is the total number of pixels in the spectrum and C is the two-dimensional cross correlation. In order to avoid oversampling the likelihood surface, we calculate the likelihood in 36 km/s steps, which is approximately half of the low-resolution NIRSPEC pixel size.


Finally, the log likelihoods as a function of brown dwarf line-of-sight velocity, $v_{\mathrm{BD}}$, at each epoch can be combined into a log likelihood as a function of the line-of-sight Keplerian orbital velocity, $K_{\mathrm{BD}}$. We take the cuts along the maximum sky emission likelihood (near 0 km/s), and convert the $v_{\mathrm{BD}}$ to $K_{\mathrm{BD}}$, assuming a circular orbit, by
\begin{equation} \label{bdvelocity}
    v_{\mathrm{BD}} = K_{\mathrm{BD}} \sin (2\pi \phi) + v_{\mathrm{sys}} - v_{\mathrm{bary}}.
\end{equation}
The systemic velocity has been measured from the white dwarf's H$\alpha$ absorption line (Table~\ref{NLTTproperties}, \citealt{longstaff2019}) and the barycentric velocity in the direction of NLTT 5306 can be calculated for the time of the observation. For Oct 17, 2019, Jan 7, 2020, and Jan 7, 2021, the barycentric velocity was 1.6 km/s, -29.6, and -29.7 km/s, respectively. The brown dwarf's orbital position, $\phi$, is calculated as,
\begin{equation}
    \phi = \frac{(T_{\mathrm{obs}}-T_o) \mod{P}}{P}
\end{equation}
where the time of inferior conjunction, $T_o$, and the orbital period, $P$, are given in Table~\ref{NLTTproperties} and $\phi$ runs from 0 to 1, with $\phi=0$ corresponding to the orbital position with the brown dwarf closest to the observer (i.e. inferior conjunction). 

As shown in Figure~\ref{Figure:orbit}, the brown dwarf orbital position varies significantly across our 5 minute exposures. For a $K_{\mathrm{BD}}$ of 400 km/s, the change in expected $v_{\mathrm{BD}}$ could be as large as 130 km/s. This is comparable to the velocity resolution of NIRSPEC in its low resolution mode ($\sim150$ km/s). As a means of accounting for this variation, we run the conversion from $v_{\mathrm{BD}}$ to $K_{\mathrm{BD}}$ 10 times, using 10 $\phi$ values for each nod, equally spaced from the start of the exposure time to the end. We then average the 10 resulting log likelihood functions. This should help to correct for the non-linear relationship between $\phi$ and $v_{\mathrm{BD}}$.





\section{Priors on $K_{\mathrm{BD}}$} \label{Section:KBDPrior}

We can leverage prior information from this system to get a sense of what to expect for the brown dwarf's Keplerian orbital velocity, $K_{\mathrm{BD}}$. The true brown dwarf orbital velocity ($2\pi a / P$), which sets a maximum limit on $K_{\mathrm{BD}}$, is 405 km/s. The line-of-sight velocity would equal this if the system were completely edge-in, with an inclination of 90$^{\circ}$. However, because the system is known to be non-transiting \citep{Steele2013}, $K_{\mathrm{BD}}$ must be less than 405 km/s. Furthermore, \citealt{Steele2013} saw no trace of either a full or even a grazing eclipse in phase-folded $i'$-band light-curves of NLTT 5306 taken with the Wide Field Camera on the Isaac Newton Telescope (INT). Given their data set-up, this is more likely explained by the system not transiting than by their missing the eclipse. The inclinations that would correspond to the minimum angles which would result in full and partial transit geometries are given by, 
\begin{equation}
i_{\mathrm{full}} = 90^{\circ} - \sin^{-1}\bigg{(}\frac{R_{\mathrm{BD}}-R_{\mathrm{WD}}}{a}\bigg{)}
\end{equation}
and 
\begin{equation}
i_{\mathrm{partial}} = 90^{\circ} - \sin^{-1}\bigg{(}\frac{R_{\mathrm{BD}}+R_{\mathrm{WD}}}{a}\bigg{)}. 
\end{equation}
Using evolutionary brown dwarf radius of $0.095 \pm 0.004$ R$_{\sun}$ \citep{Steele2013}, we find that any inclination above $81.9 \pm 0.4^{\circ}$ would correspond to a full transit and any above $78.7 \pm 0.4^{\circ}$ would correspond to a partial transit. These inclinations can be converted to Keplerian orbital velocities by,
\begin{equation}
K_{\mathrm{BD}} = \frac{M_{\mathrm{WD}}K_{\mathrm{WD}}\sin(i)}{M_{\mathrm{BD}}\sin (i)},
\end{equation}
and give upper limits of $381 \pm 45$ (to exclude a full eclipse) and $378 \pm 45$ km/s (to exclude a partial eclipse). Using the Roche lobe radius of $0.12 \pm 0.02$ R$_{\sun}$, the full transit would extend down to $i=79 \pm 2^{\circ}$ or up to $K_{\mathrm{BD}} = 378 \pm 45$ km/s and the partial transit would go down to $i=76 \pm 2^{\circ}$ or up to $K_{\mathrm{BD}} = 374 \pm 44$ km/s. Assuming the brown dwarf has a radius between the evolutionary limit and Roche lobe radius, and that this system does not show even a partial eclipse, the upper limit of $K_{\mathrm{BD}}$ is between $378 \pm 45$ and $374 \pm 44$ km/s. 


We can approximate a lower limit on $K_{\mathrm{BD}}$ from the expected mass of the brown dwarf. \citealt{Steele2013} estimated that NLTT 5306B has a spectral type of L4-L7 based on two H$_2$O indices defined by \citealt{Burgasser2002} measured from a near-infrared X-shooter spectrum of NLTT 5306B. \citealt{Casewell2020nltt5306} further refined the brown dwarf's spectral type to L5 based on a SpeX $JHK$ spectrum. As NLTT 5306B is confidently a brown dwarf, as opposed to a star, we can set an upper limit on its mass at the hydrogen burning limit of $\sim75$ M$_{\mathrm{Jup}}$, below which electron degeneracy pressure prevents the object's core from reaching the temperatures needed for nuclear fusion \citep{Hayashi1963,Kumar1963}. An upper limit on the brown dwarf mass of 75 M$_{\mathrm{Jup}}$ would correspond to a lower limit on $K_{\mathrm{BD}}=K_{\mathrm{WD}}M_{\mathrm{WD}}/M_{\mathrm{BD}}$ of $288 \pm 30$ km/s.

From NLTT 5306B's expected mass and lack of even a partial eclipse, then, we can deduce $K_{\mathrm{BD}}$ should be in between about 288 and 378 km/s. This 90 km/s range is less than the velocity resolution of our NIRSPEC data ($\sim$150 km/s). 

\section{Results}  \label{Section:NLTTresults}

We cross correlate the NIRSPEC data with each of the Sonora and irradiated models described in Section~\ref{sonoramodels} and then compare their probability values. As the prior constraints on $K_{\mathrm{BD}}$ are stronger than we could make with our data, we compare the average of the log likelihood values between 288 and 378 km/s. We convert the log likelihoods calculated by Equation~\ref{logLequation} to probabilities to compare them. 

\begin{figure*}
    \centering
    \noindent\includegraphics[width=42pc]{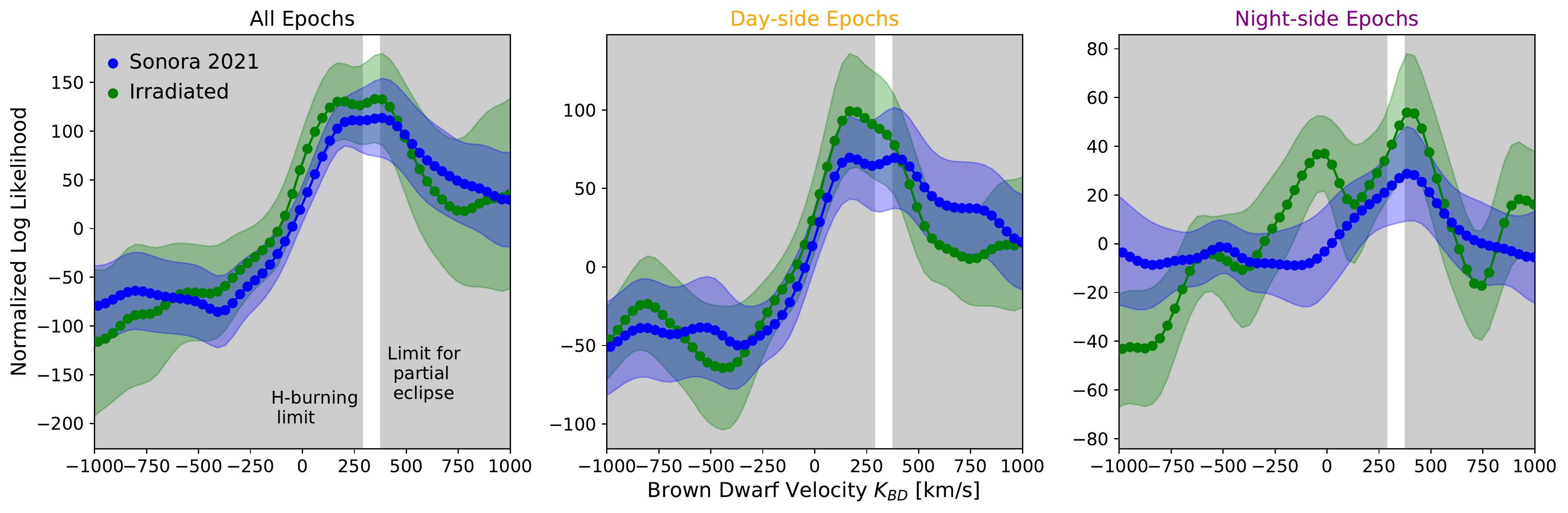}
    \caption{Normalized log likelihood functions generated from the cross correlation between the conditional best fitting model from each model grid to our data, shown with jack-knife error bars. The three panels represent fits to the full data set, to only the day-side epochs, and to only the night-side epochs. The blue curves come from Sonora 2021 models and the green curves are from irradiated models. The vertical white range shows the prior on $K_{\mathrm{BD}}$ given that the companion is a brown dwarf and does not show even a grazing transit.      }
    \label{Figure:KpcurveDayandNight}
\end{figure*}

Figure~\ref{Figure:KpcurveDayandNight} shows the results from the conditionally best fitting Sonora and irradiated models. The left-most panel shows an example fit from the Sonora grid and from the irradiated grid to the full 44 epoch data set, each shown with jackknife errorbars. 

We also fit subsets of the data containing only day-side facing and only night-side facing epochs. Day-side epochs have $\phi$ values between 0.25 and 0.75, shown in orange in Figure~\ref{Figure:orbit}, and night-side epochs have $\phi$ values between 0.75 and 0.25, shown in purple. In total, we had 23 day-side epochs and 18 night-side epochs. Three epochs crossed quadrature ($\phi=0.25,0.75$) during their 5-min exposures meaning they would have shown roughly 50\% brown dwarf day-side and 50\% night-side. We discarded these epochs from the day/night-side analysis.

The center and right-most panels of Figure~\ref{Figure:KpcurveDayandNight} show example fits to only the day-side epochs and only the night-side epochs, respectively. When fitting the full data set, the best fitting Sonora model result in a likelihood peak about $4\times$ over the baseline, whereas when fitting only the day-side epochs, we see a peak around $3\times$ over the baseline, and with only night-side epochs, a peak arises around $2\times$ over the baseline. The irradiated models show roughly the same level fits to the full data set and day-side only subset, but shows significantly more noise than the Sonora model against the night-side only subset.

Corner plots showing the relative likelihood across the full model grids are shown in Figures~\ref{Figure:Sonoracornerplots} (Sonora grid) and ~\ref{Figure:Irradiatedcornerplot} (irradiated grid). The left-most corner plots fit the full data set, and the center and right-most corner plots fit the day- and night-side only subsets, respectively. 


\subsection{Sonora Analysis}

\begin{figure*}
    \centering
    \noindent\includegraphics[width=42pc]{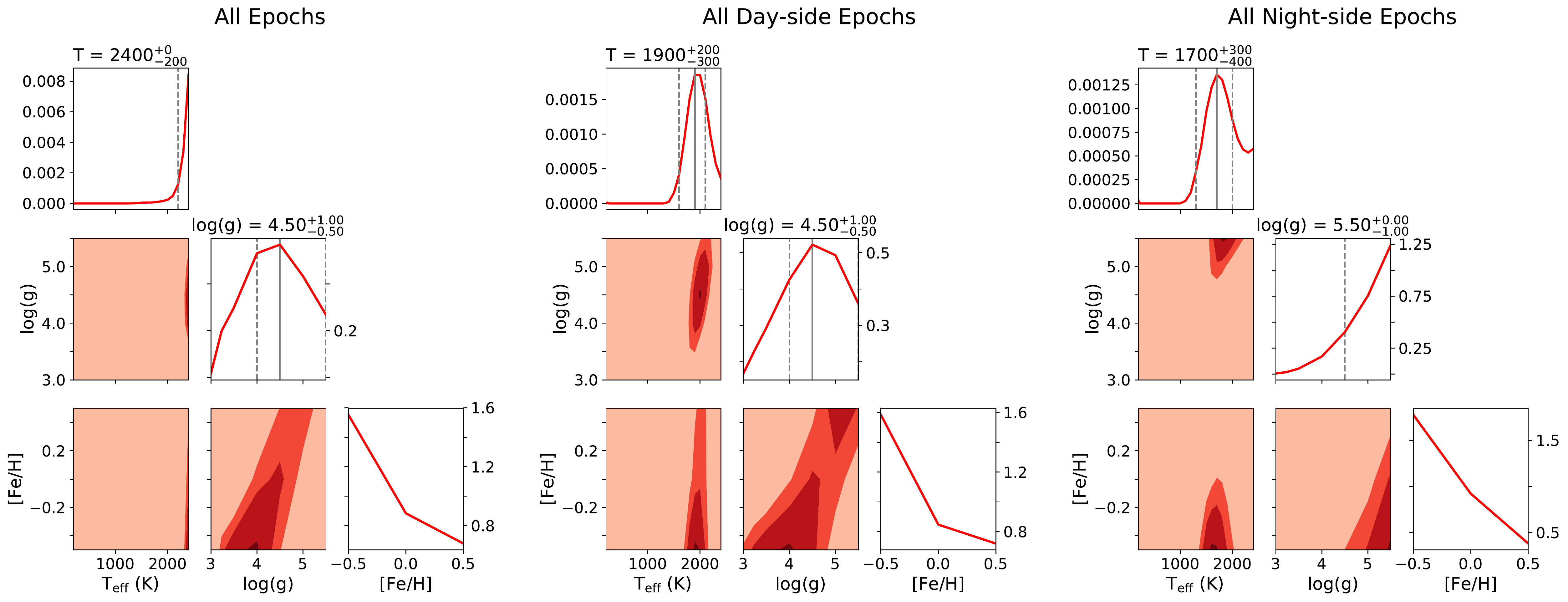}
    \caption{Results of Sonora model fits. The three corner plots show Sonora fits to all of the epochs, the NLTT 5306B day-side epochs, and the NLTT 5306B night-side epochs. The contour plots show higher likelihood with darker colors, and the line plots the marginalized results with 68\% confidence intervals. Contours are at 50, 68, 95\%. The log likelihood value used to compare each model is the average of the log likelihoods between $K_{\mathrm{BD}}$ of 288 and 378 km/s.    }
    \label{Figure:Sonoracornerplots}
\end{figure*}

Comparisons of the full suite of Sonora 2021 models to our data sets -- all epochs, day-side epochs, and night-side epochs -- are shown in Figure~\ref{Figure:Sonoracornerplots}. The marginalized, with 68\% confidence intervals, and conditional best fitting models are reported in Table~\ref{Table:Sonorapicks} and graphed in Figure~\ref{Figure:KpcurveDayandNight}.

We first consider the effective temperatures favored by our data. The day-side epoch subset of our data prefers models with effective temperatures of 2000 K, or, when marginalized, $1900^{+200}_{-300}$ K. The night-side epoch subset, on the other hand, prefers the 1800 K model, or when marginalized, $1700^{+300}_{-400}$ K. There is substantial overlap in the marginalized effective temperature likelihoods between the day- and night-side epoch subsets, implying that there is minimal temperature difference. This is consistent with the conclusions \citet{Casewell2020nltt5306} drew from the day- and night-side brightness temperatures of NLTT 5306B in the $Spitzer$ wavebands. 

Figure~\ref{Figure:BrightnessTemps} shows the relationship between our effective temperature measurements and the brightness temperatures from \citet{Casewell2020nltt5306}. The brightness temperatures shown with squares were calculated under the assumption that the brown dwarf has a radius predicted by evolutionary models, while the temperatures shown with circules assumed a Roche lobe radius. If the brown dwarf is uniformly inflated, as suggested by \citet{Casewell2020nltt5306}, its brightness temperatures should lie in between the two predictions. Our night-side effective temperature measurement is quite consistent with the brightness temperatures, while our day-side temperature is a bit hotter.

Our day-side epochs prefer somewhat lower gravity models than our night-side epochs. The day-side epochs prefer models with a $\log(g)=4.5^{+1}_{-0.5}$, while the night-side models prefer $\log(g)=5.5_{-1.0}$. The later is a lower limit rather than a true measurement because $\log(g)=5.5$ lies at the edge of the model grid. The two-dimensional likelihood surfaces show a more significant difference in gravities than in effective temperatures. \citealt{Casewell2020nltt5306} showed that an intermediate gravity ($\log(g)\sim 4.8$) template better fit a $R\sim120$ $JHK$ SpeX spectrum of NLTT 5306B than a field gravity ($\log(g)\sim 5.2$) template, although this spectrum was observed over roughly half an orbit of the system, which would make it impossible to detect phase variation in surface gravity.

Finally, whether we consider the day-side, night-side, or all 44 epochs, our data prefer the low metallicity, [Fe/H] $=-0.5$, models. This is consistent with what we would expect for an object with the $> 5$ Gyr system age of NLTT 5306 \citep{Steele2013}. We do see, though, especially in the corner plots of the full data set and day-side only subset, a degeneracy between high surface gravity and low metallicity models that is likely arises from these parameters' common effect on spectral morphology (see Section~\ref{sonoramodels}). 


\begin{deluxetable}{lccc} 
\tablewidth{0pt}
\def\arraystretch{1}
\tablecaption{Best Fitting Models}
\tablehead{Property & All epochs & Day-side  & Night-side  }
\startdata
\multicolumn{4}{c}{\textbf{Sonora 2021 Model Grid}} \\
\multicolumn{4}{l}{\textbf{Marginalized}} \\
$T_{\mathrm{eff}}$ (K) & 2400$^{+0}_{-200}$ & $1900^{+200}_{-300}$ & $1700^{+300}_{-400}$ \\
$\log(g)$  & 4.5$^{+1.0}_{-0.5}$ & 4.5$^{+1}_{-0.5}$ & 5.5$^{+0}_{-1.0}$ \\
$[$Fe/H$]$  & -- & --  & -- \\
   \multicolumn{4}{l}{\textbf{Conditional}} \\
$T_{\mathrm{eff}}$ (K) & 2400 & 2000 & 1800  \\
$\log(g)$   & 4.0 & 4.0 & 5.5 \\
$[$Fe/H$]$  & -0.5 & -0.5 & -0.5 \\
\multicolumn{4}{c}{\textbf{Irradiated Model Grid}} \\
\multicolumn{4}{l}{\textbf{Conditional}} \\
$T_{\mathrm{int}}$ (K) & 2000 & 2000 & 2000  \\
$\log(g)$   & 4.5 & 5.0 & 5.0 \\
$[$Fe/H$]$  & 0.5 & -0.5 & -0.5 \\
$f$     &    0.125   &  0.5     &  0.25     
\enddata
\label{Table:Sonorapicks}
\end{deluxetable}


\subsection{Irradiated Model Analysis}

\begin{figure*}
    \centering
    \noindent\includegraphics[width=42pc]{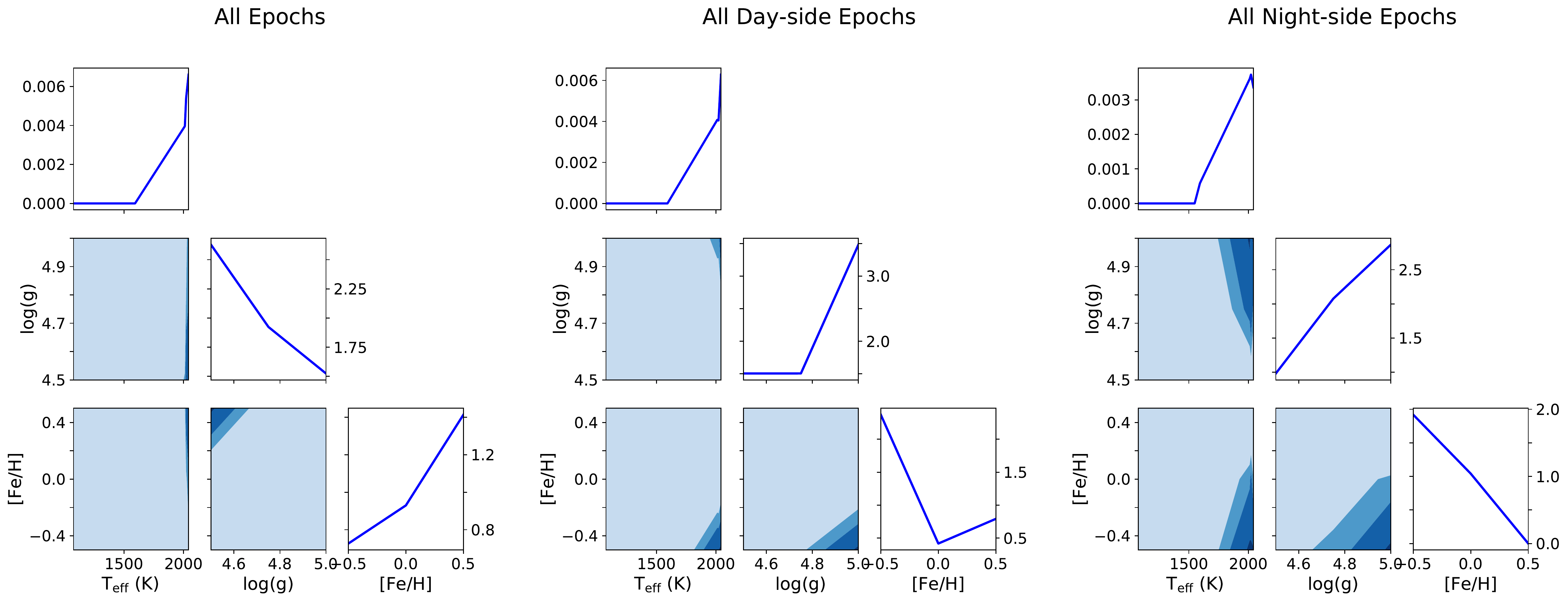}
    \caption{Same as Figure~\ref{Figure:Sonoracornerplots}, but showing the results of the irradiated model fits. We convert the model internal temperature and heat redistribution parameters to brown dwarf effective temperatures using Equation~\ref{Equation:Tintf_Teff} and assuming a Bond albedo of 0. }
    \label{Figure:Irradiatedcornerplot}
\end{figure*}

Figure~\ref{Figure:Irradiatedcornerplot} shows how the irradiated model grid fits the full suite of data, as well as the day-side and night-side only subsets. To plot these results in an analogous fashion to the Sonora results, we convert the brown dwarf internal temperature and heat redistribution parameters to a brown dwarf effective temperature as described in Section~\ref{sonoramodels}. 

As mentioned, there is a degeneracy between $f$ and $A_{\mathrm{BD}}$, such that the choice of $A_{\mathrm{BD}}$ makes little difference to our results here. Setting $A_{\mathrm{BD}}$ to 0 versus 1 does not affect the probability values in any way; it slightly alters the effective temperatures from the best fitting internal temperatures depending on the magnitude of the best fitting heat redistribution parameters. 

All subsets prefer an internal temperature of 2000 K, though disagree on the preferred heat redistribution parameter (Table~\ref{Table:Sonorapicks}), leading to slightly different effective temperatures. The full data set selects $T_{\mathrm{eff}} = 2011$ K, while the day- and night-side subsets select 2042 and 2021 K, respectively. 

Interestingly as well, the day- and night-side subsets agree on a low metallicity ([Fe/H] $=-0.5$), higher gravity ($\log(g) = 5.0$) model while the full data set selects the opposite: a high metallicity ([Fe/H] $=0.5$), lower gravity ($\log(g)=4.5$) model. It is surprising that with the day- and night-side subsets together making up 93\% of the full data set that we see this disagreement. One explanation may be that the night-side detection is not as robust as the day-side or full data set detections or as the Sonora detections. As can be seen in Figure~\ref{Figure:KpcurveDayandNight}, the irradiated model night-side detection, while giving a higher likelihood within the desired velocity range, also shows significantly more noise structure than the Sonora night-side detection. 





\subsection{Comparison of Model Grids}

\begin{figure}
    \centering
    \noindent\includegraphics[width=21pc]{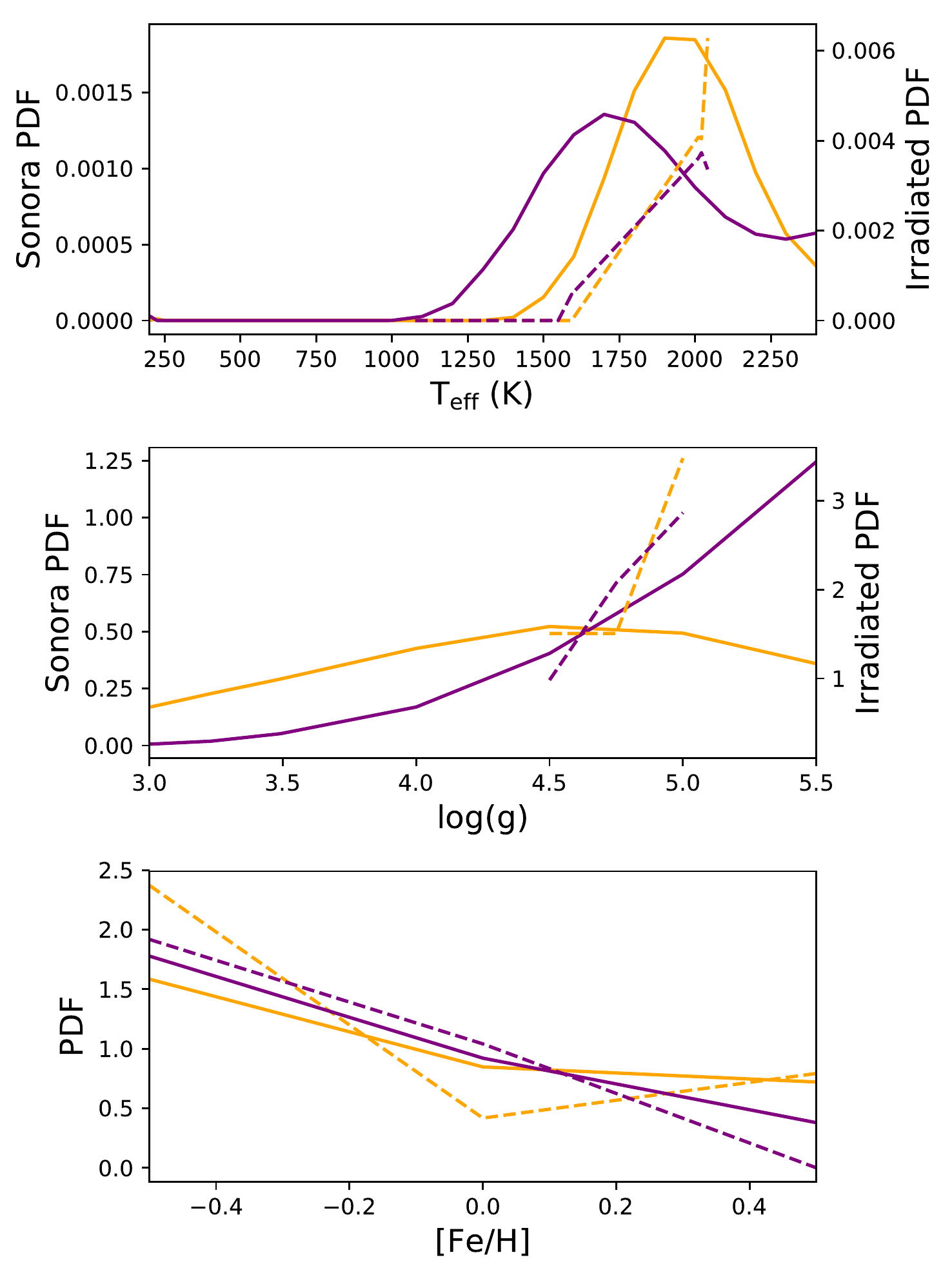}
    \caption{ Comparison between the Sonora (solid curves) and irradiated (dashed curves) model fits to our data. The orange curves were fit to day-side epochs and the purple curves were fit to night-side epochs. Since the two model grids cover very different ranges of effective temperatures and surface gravities, we plot their normalized probability distribution functions on separate y-axes.  }
    \label{Figure:comparemodels}
\end{figure}

Figure~\ref{Figure:comparemodels} compares the normalized probabilities from the Sonora and irradiated model grid fits. The solid curves represent the Sonora distributions and the dashed curves represent the irradiated distributions. The orange curves come from the day-side only subset of data while the purple curves come from the night-side only data. Because the Sonora and irradiated model grids offer different ranges of effective temperatures and surface gravities, we put their normalized posterior distribution functions on separate y-axes. 

There is very good agreement in the metallicity results. The irradiated model grid tends toward higher effective temperatures and gravities than the Sonora grid, but the irradiated grid offers quite a small range of each parameter and we cannot resolve the shape of the probability distribution function as well as with the Sonora models. Future advancements in the modeling of irradiated objects like NLTT 5306B, including, for example, more detailed grids, could further our interpretation of the data.

\section{Discussion}

\subsection{Effective Temperature and Gravity }
\begin{figure}
    \centering
    \noindent\includegraphics[width=21pc]{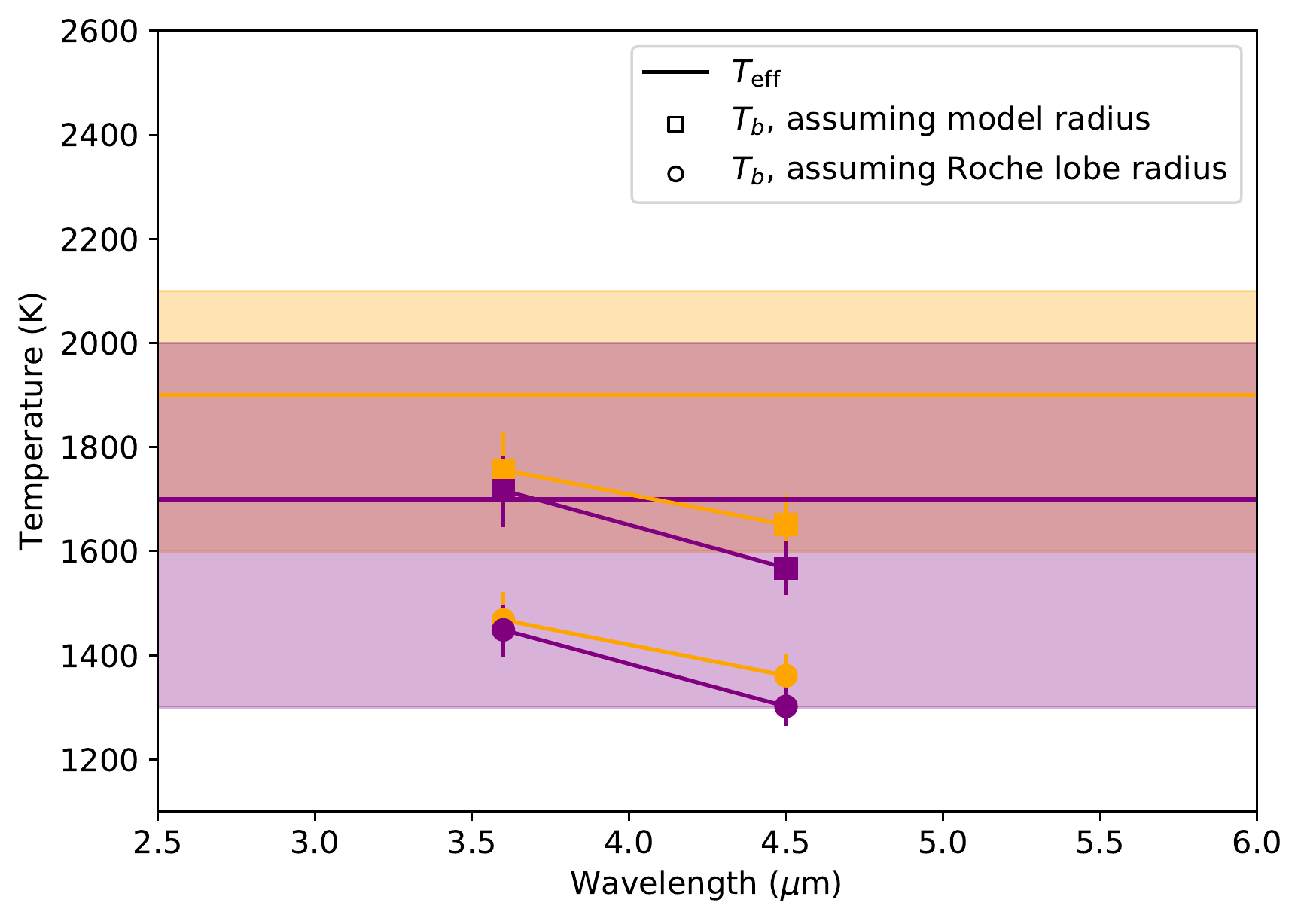}
    \caption{  NLTT 5306 B's wavelength dependent brightness temperatures (points, from \citealt{Casewell2020nltt5306}) and effective temperatures (horizontal lines). Day-side measurements are in orange and night-side measurements are in purple. The brightness temperatures shown with squares were calculated assuming the evolutionary model radius, and the crosses assumed the Roche lobe radius.       }
    \label{Figure:BrightnessTemps}
\end{figure}

Our Sonora 2021 analysis presented some interesting results. As illustrated in Figure~\ref{Figure:BrightnessTemps}, the day- and night-side effective temperatures are fairly consistent with each other, but the day-side temperature is a bit hotter. The night-side temperature was also very consistent with brightness temperature estimates from \citealt{Casewell2020nltt5306}. We also saw that the day-side favored lower gravity models than the night-side, though overall the night-side detection was not as strong as the day-side detection. This difference in detection strength is expected since, at these effective temperatures, our $\sim$1 $\mu$m data fall on the Planck side of the black body function where small drops in $T_{\mathrm{eff}}$ dramatically affect flux.

While our night-side effective temperature is very consistent with the brightness temperatures, if we consider the slightly raised day-side effective temperature to be significant, one explanation for it could be the presence of a hot spot. As compared to hot Jupiters, brown dwarfs orbiting white dwarfs are extremely rapid rotators, by means of their smaller orbital (and, thus, rotation) periods. This faster rotation can lead to smaller eastward-shifted hot spots, but more significant westward-shifted hot regions which arise from off-equatorial Rossby gyres \citep{Tan2020}. As \citealt{Zhou2021} describe, while brightness temperatures measured from low resolution data are hemispherically averaged quantities, effective temperatures, which come through higher resolution spectral fitting, would be more sensitive to hot spots. As hot spots dominate spectral emission, even if they do not cover the majority of the visible surface area, they can bias effective temperature measurements to higher values than the band-averaged brightness temperatures. The day-side effective temperature being raised over both the night-side effective temperature and the brightness temperatures could be explained if the hot spot was only visible on the day-side of the brown dwarf. The day-side of the brown dwarf is the side facing the white dwarf. Perhaps then, a day-side hot spot could be related to the apparent accretion onto the white dwarf \citep{longstaff2019}. 

Our lower day-side gravity measurement may fit in to this picture. If the white dwarf were pulling some matter from the brown dwarf surface facing it (which we witness during the brown dwarf's day-side), we may expect to see a hotter and lower gravity region. 

This would imply, however, that the brown dwarf surface is distorted. \citealt{Casewell2020nltt5306} have shown that distortion by its interaction with the white dwarf is unlikely. Using the mass ratio, separation, and assumed radius of the brown dwarf (0.095 R$_{\sun}$), they calculated that the tidal distortion due to the white dwarf and the tidally locked rotation, is only 2.5\%. They predicted that the majority of this distortion was due to the rotation rate. While distortion could account for 2.5\% difference between the equatorial and polar radii, the difference between NLTT 5306B's model radius and the radius of an intermediate gravity brown dwarf of the same mass is 22\%. While with its model radius of 0.095 R$_{\sun}$, NLTT 5306B is filling $\approx80\%$ of its Roche lobe \citep{longstaff2019}, if it were inflated by 22\% to a radius of 0.11 R$_{\sun}$, NLTT 5306B would fill nearly its full ($\approx96\%$) Roche lobe. From these calculations, \citealt{Casewell2020nltt5306} predicted that the lower gravity signatures they saw in SpeX low resolution data were not likely due to distortion and more likely represented a brown dwarf which was uniformly inflated. 

Our results do show day- and night-side differences in preferred surface gravity. If distortion is an unlikely cause of these differences, it may be interesting to look back on how our measurements were made. We compare the different models based on the average of their likelihoods fit to the data between a $K_{\mathrm{BD}}$ of 288 and 378 km/s. As described above, this velocity range prior was found on one end from the mass cut-off between stars and brown dwarfs and on the other end from the lack of a partial eclipse. For this system, the white dwarf has a line-of-sight velocity of -46.8 km/s \citep{longstaff2019}. Material streaming from the brown dwarf to the white dwarf should start with the brown dwarf's velocity and gradually transition to the white dwarf's velocity. While the velocity range we considered is less than the velocity resolution of our NIRSPEC data, the 90 km/s range could still include the brown dwarf velocity component at a higher end of the velocity range and some of the material en route to the white dwarf towards the lower end of the range. 

The non-hydrostatic material streaming between the objects could bias our measurement of the ``bulk" day-side surface gravity to lower gravities. During the brown dwarf's night-side, however, the brown dwarf's cross section would likely cover the majority of the material streaming between the two and not show this same bias. That our night-side surface gravity is more consistent with the evolutionary radius matches these predictions.

While above we postulated how a hot spot could increase the day-side effective temperature, the day- and night-side effective temperatures share significant overlap. Their 68\% confidence intervals do overlap between 1600 and 2000 K. In this case, a hot spot may not be needed to explain our results. This range of overlapping temperatures is closer to the brightness temperatures calculated assuming the evolutionary model radius, or higher gravity brown dwarf. Further, the overlapping temperature range is only marginally hotter than the 1581 K \citet{Filippazzo2015} would predict for an L5 field age dwarf from their sixth order polynomial fit to 124 field age objects.


From this perspective, we might envision a non-distorted non-inflated brown dwarf with an evolutionary radius. Its day- and night-side effective temperatures are consistent, as was also seen with its brightness temperature. The brown dwarf night-side, while not overall strongly detected, prefers higher gravity values, and the effective temperatures are somewhat more consistent with the brightness temperatures derived assuming an evolutionary radius (see Fig.~\ref{Figure:BrightnessTemps}). In this picture, the lower gravity component we seen in the day-side may rise completely from material streaming off of the brown dwarf, and so may be considered distinct from the brown dwarf itself. 

However, \citealt{Casewell2020nltt5306} did show evidence that NLTT 5306B has an intermediate gravity surface rather than the evolutionarily expected high gravity in a low resolution SpeX $JHK$ spectrum. Further, WD1032B, a brown dwarf that eclipses a similarly cool white dwarf to NLTT 5306A and receives $\sim$1.5 times the irradiation of NLTT 5306B, is inflated \citep{Casewell2020WD1032}, as determined by its radius measurement. Could the intermediate gravity features seen in the SpeX spectrum of NLTT 5306B be consistent with a noninflated brown dwarf and lower gravity stream of material rather than a uniformly inflated object? Two pieces of information may be relevant. First, the SpeX data were taken over roughly half of NLTT 5306B's orbit, which would make it impossible to detect phase variation in surface gravity. Second, SpeX data have a resolution of $R\sim 120$, corresponding to a velocity resolution of approximately 2500 km/s. This resolution would convolve all of the velocity components of the system, from the white dwarf to the brown dwarf, together. While the white dwarf has a very different spectral shape than the brown dwarf, and could be distinguished in this way, the material streaming between the two may more closely resemble the brown dwarf spectroscopically. If so, and if the streaming material were indeed lower gravity than the brown dwarf, the intermediate gravity features \citealt{Casewell2020nltt5306} presented could be from a linear combination of the higher gravity brown dwarf and lower gravity material.  

A final possibility could come from our observational set up. As described in Section~\ref{Section:NLTTobservations}, our 5 minute observations allow some change in the brown dwarf velocity. We can predict the expected changes in velocities across the day-side versus night-side epochs. Assuming the maximum $K_{\mathrm{BD}} = 378$ km/s, the velocity change over day-side epochs varies from 6 to 164 km/s, with a mean of 87 km/s, while the night-side variation ranges from 9 to 126 km/s, with a mean of 78 km/s. With a higher mean, the day-side epochs show slightly more change in velocity. This could act to broaden out spectral features in these epochs. \citealt{Allers2013} described how low-gravity brown dwarfs show weaker FeH bands, Na I lines, and K I lines than do field gravity brown dwarfs. The velocity effects built in to our data set could work to broaden out these spectral features in our day-side data more so than in our night-side data, making them appear weaker, and thus leading to a preference for lower gravity models.


\subsection{Metallicity and Magnetism}

Our metallicity constraints provide a final compelling clue. We find that our data, whether considering all 44, only day-side, or only night-side epochs, prefer the [Fe/H] $=-0.5$ Sonora models. The day- and night-side subsets of data also prefer low metallicity irradiated models. This is consistent with what we would expect from a thick disk object, of a considerable age ($>5$ Gyr, \citealt{Steele2013}). Yet, in pondering potential mechanisms for the brown dwarf's inflation, \citealt{Casewell2020nltt5306} cited a high metallicity, cloudy atmosphere. \citealt{Burrows2011} showed that the difference in radii between a clear, [Fe/H] $=-0.5$ brown dwarf and a cloudy, [Fe/H] $=0.5$ one could be $\sim0.25 R_{\mathrm{Jup}}$ at early ages and $\sim0.1 R_{\mathrm{Jup}}$ at late ages. By these estimates, to achieve the $\sim0.2 R_{\mathrm{Jup}}$ inflation \citealt{Casewell2020nltt5306} estimated from the SpeX data of NLTT 5306 B, the brown dwarf would need a metallicity near the upper end of the range ([Fe/H] $= 0.5$) as well as a cloudy atmosphere. Our Sonora results show that this is unlikely. Two conclusions can be drawn from this finding. As metallicity is not likely responsible for the inflation of NLTT 5306 B, either the brown dwarf is not inflated or if it is, its inflation must be a result of some other process.  

We provide direct evidence that the suspected inflation of NLTT 5306 B is not due to a high metallicity, cloudy atmosphere. This finding amplifies the evidence that magnetic evidence is at play, as was discussed in depth in \citealt{Casewell2020nltt5306}. High resolution spectropolarimetric measurements could confirm this theory. We can set a prior on the strength of NLTT 5306 A's magnetic field from earlier findings. \citealt{longstaff2019} saw no evidence of an infrared excess characteristic of an accretion disk accompanying the H$\alpha$ emission feature on NLTT 5306 A's surface that led to the conclusion it was accreting mass from its brown dwarf. White dwarfs called ``polars" have strong enough magnetic fields that they can funnel mass along field lines, preventing the formation of an accretion disk. With the estimated accretion rate, \citealt{longstaff2019} calculated NLTT 5306 A's magnetic field must be at least $0.45 \pm 0.02$ kG to prevent an accretion disk forming. On the other end, there were no signs of either Zeeman splitting in NLTT 5306 A's Balmer lines or cyclotron humps in its XSHOOTER \textit{JHK} spectrum \citep{longstaff2019}. Typical white dwarf spectra are significantly pressure broadened, to a full width at half maximum (FWHM) of more than 1 \AA, enough to blur out Zeeman splitting from fields up to $\sim 30 - 50$ kG \citep{Landstreet2017}. Even stronger fields would be needed for cyclotron emission to arise in the $J$ band; the cyclotron fundamental is detectable at optical to near-IR wavelengths for fields of strength $\sim 10^5$ kG \citep{Ferrario2020}. It follows that NLTT 5306 A's magnetic field must lie in the range from 0.45 to $\sim 30 - 50$ kG. 

For magnetic activity to explain the inflation of NLTT 5306B, NLTT 5306 A should show a stronger field than the white dwarfs hosting non-inflated brown dwarfs. Recall that NLTT 5306 A ($T_{\mathrm{eff}} = 7756$ K) and WD1032+011A ($T_{\mathrm{eff}} = 9950$ K), the two white dwarfs hosting inflated brown dwarfs, are cooler than SDSS J1205-0242 ($T_{\mathrm{eff}} = 23680$ K) and SDSS J1411+2009 ($T_{\mathrm{eff}} = 13000$ K), the two hosting non-inflated brown dwarfs. In fact, observations suggest that a higher fraction of white dwarfs with lower effective temperatures have strong magnetic fields. \citealt{Hollands2015} found a $13 \pm 4\%$ incidence of magnetic activity in a sample of DZ white dwarfs with $T_{\mathrm{eff}} < 9000$ K, much higher than the incidence among young, hot DA white dwarfs. This could be because, at these low temperatures, white dwarfs are cool enough to be (at least partially) crystallized, which has been proposed as one method of generating magnetic fields \citep[e.g.][]{Isern2017}.   

While NLTT 5306 is a very faint source ($g' = 17.03$, \citealt{Steele2013}), \citealt{Bagnulo2018} recently published a survey of the weakest detectable magnetic fields in white dwarfs and concluded that both the European Southern Observatory's Very Large Telescope's low-resolution spectropolarimeter, FORS2, and the \textit{William Herschel} Telescope's mid-resolution spectropolarimeter, ISIS, could search for mean longitudinal fields $\langle B_Z \rangle \sim 1$ kG in $V_{\mathrm{mag}} \lesssim 14$ DA stars. The extension to NLTT 5306 A could not only answer questions about the unique forces acting on NLTT 5306 B, covering nearly the full magnetic field strength prior, but would also expand the population of white dwarfs we can target to fill in gaps about how white dwarf magnetism scales with mass, age, and rotation, an issue described by \citealt{Bagnulo2018}.

Heating of NLTT 5306B's deep atmosphere by vertical advection of potential temperature, as was described in \citealt{Sainsbury2021}, can also not be ruled out as a mechanism responsible for radius inflation. Next-generation, fully radiative 3D global circulation models (GCMs) could test this theory.



\section{Conclusions}

Ultimately, while we have some understanding of NLTT 5306B, there is still much to learn. The results from our Sonora study could be consistent with either a hot and distorted spot on the brown dwarf day-side or with a evolutionary radius brown dwarf with no significant day- to night-side temperature difference, but some traces of the lower gravity detached material streaming to the white dwarf. Still, we do not know precisely \textit{why} this system is interacting. 

We do know that gravitational distortion is insufficient to explain the perceived inflation of NLTT 5306B. And, we know that if there is inflation, it is certainly not to the Roche lobe level, from the upper limit set on the mass accretion rate in the system \citep{longstaff2019}. A high metallicity, cloudy atmosphere is not likely responsible for this suspected inflation. Additionally, the atmosphere is not so hot as to be boiled off. Other, hotter brown dwarfs in equivalent systems show no signs of interaction, such as SDSS J141126.20+200911.1 and SDSS J120515.80024222.6 \citep{Casewell2020WD1032}. Magnetic activity from the white dwarf could be inflating the brown dwarf, as is seen in M dwarfs, and would be a ripe source for future investigations.


\acknowledgments{ 
We thank Shreyas Vissapragada, Jessica Spake, and Antonija Oklop\u{c}i\'{c} for interesting conversations along the way. We also thank our anonymous reviewer for their helpful feedback on this manuscript.

The data presented herein were obtained at the W. M. Keck Observatory, which is operated as a scientific partnership among the California Institute of Technology, the University of California and the National Aeronautics and Space Administration. The Observatory was made possible by the generous financial support of the W. M. Keck Foundation. The authors wish to recognize and acknowledge the very significant cultural role and reverence that the summit of Maunakea has always had within the indigenous Hawaiian community.  We are most fortunate to have the opportunity to conduct observations from this mountain.

The irradiated atmosphere models were calculated as part of program HST-AR-16142. Support for HST-AR-16142 was provided by NASA through a grant from the Space Telescope Science Institute, which is operated by the Association of Universities for Research in Astronomy, Incorporated, under NASA contract NAS5-26555. This work also benefited from support from the NASA Exoplanet Research Program (grant NNX16AI14G, G.A. Blake P.I.).}

{\footnotesize
\bibliography{main}}
\bibliographystyle{ApJ}

\end{document}